\documentclass[prb,twocolumn,superscriptaddress,showpacs]{revtex4}
\usepackage{amsmath}
\usepackage{amssymb}
\usepackage{epsfig}
\usepackage{graphicx}
\usepackage{wasysym}
\usepackage{subfigure}
\usepackage{psfrag}
\newcommand \be{\begin{equation}}
\newcommand \ee{\end{equation}}
\newcommand \bes{\begin{equation*}} 
\newcommand \ees{\end{equation*}}
\newcommand \bea{\begin{eqnarray}}
\newcommand \eea{\end{eqnarray}}
\newcommand \beas{\begin{eqnarray*}} 
\newcommand \eeas{\end{eqnarray*}}
\newcommand \bfg{\begin{figure}}
\newcommand \efg{\end{figure}}
\newcommand \bfgs{\begin{figure*}} 
\newcommand \efgs{\end{figure*}}
\newcommand \bwt{\begin{widetext}} 
\newcommand \ewt{\end{widetext}}

\begin{document}
\title{Kinetic Magnetism and Orbital Order
in Iron Telluride}

\author{Ari M. Turner}
\affiliation{Department of Physics, University of California at
Berkeley, Berkeley, CA 94720}

\author{Fa Wang}
\author{Ashvin Vishwanath}
\affiliation{Department of Physics, University of California at
Berkeley, Berkeley, CA 94720} \affiliation{Materials Sciences
Division, Lawrence Berkeley National Laboratory, Berkeley, CA 94720}

\date{\today}

\begin{abstract}
Iron telluride (FeTe), a relative of the iron based high temperature superconductors, displays unusual magnetic order and structural transitions. Here we explore the idea that strong correlations may play an important role in these materials. We argue that the unusual orders
observed in FeTe can be understood from a picture of correlated
local moments with orbital degeneracy, coupled to a small density of itinerant electrons. A component of the structural transition is
attributed to orbital, rather than magnetic ordering, introducing a
strongly anisotropic character to the system along the diagonal
directions of the iron lattice. Double exchange interactions couple
the diagonal chains leading to the observed ordering wavevector.
The incommensurate order in samples with excess iron arises from electron doping in this scenario. The strong anisotropy of physical properties in the ordered phase should be detectable by transport in single domains. Predictions for ARPES, inelastic neutron scattering and hole/electron doping studies are also made.
\end{abstract}

 \maketitle
\section{Introduction}
The discovery of high temperature superconductivity in a class of
iron based materials \cite{Kamihara} has opened a new route to high temperature superconductivity besides the ones operating in the copper oxide materials. Following the initial discovery in LaOFeAs (1111
materials), a number of new classes of materials were discovered
that shared similar properties, notably the 122 materials (like
BaFe$_2$As$_2$). In these systems, (collectively referred to as the
FeAs materials) the undoped compound is a metallic SDW system, with
ordering wavevector $(\pi,0)$, which on doping leads to a
superconducting state. The magnetism in these materials is believed
to arise from Fermi surface nesting, given the presence of an
electron and hole pocket separated by $(\pi,0)$ in the LDA
band structure calculations of these materials. In fact, theoretical
studies had predicted this ordering before it was confirmed in
neutron scattering experiments. Moreover, the ordering moment is
typically quite small, eg. of order $0.3\mu B$ in LaOFeAs, and the
absence of a Curie-Weiss form of magnetic susceptibility above the
ordering temperature have been invoked as evidence for the itinerant
character of the magnetism. Finally, signatures of an
excitation gap appear in optical conductivity experiments, on
cooling through the SDW transition.\cite{NLWang1}

\begin{figure}
\includegraphics[width=.45\textwidth]{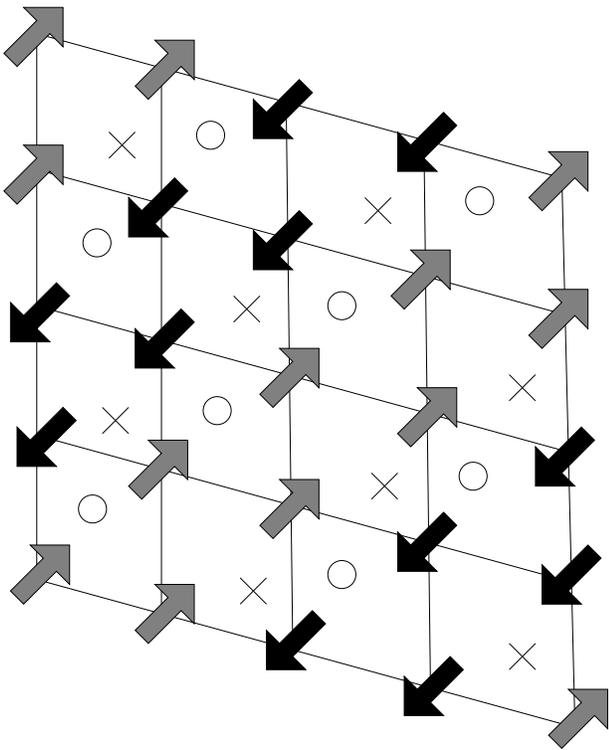}
\caption{\label{fig:rhombi} The commensurate $(\frac{\pi}{2},\frac{\pi}{2})$
spin ordering pattern of Fe$_{1+x}$Te$_1$, close to $x=0$. The spins reside on the iron atoms,
which form a square lattice. The tellurium atoms alternate above (x's) and below (o's) the iron
plane.  When the spins order, the square lattice of irons distorts
into an approximately rhombic lattice.  We assume that the monoclinic
distortion is a consequence of the spin ordering.
The tellurium atoms move towards (or away from) the exceptional spin in each parallelogram.  Thus, the planes of tellurium atoms shift in opposite directions,
producing the monoclinic distortion.}
\end{figure}

An important recent development has been the discovery of a new
class of materials FeSe and FeTe, which share the square lattice Fe
structure of the FeAs materials and are believed to be closely
related. Indeed, superconductivity has been observed in FeSe even in
the absence of doping at 8$K$, rising to 37$K$ on application of
hydrostatic pressure. The chemical simplicity of these materials, as
well as the absence of a pnictide group element, may offer valuable
clues to isolating the physics of the iron based high
temperature superconductors. One notable difference from the FeAs
materials though, is in the magnetism. While FeSe is non-magnetic
even at stoichiometry, the FeTe materials are magnetically ordered
metals, but with a more complicated kind of order than seen in
FeAs, shown in Fig. \ref{fig:rhombi}.
The ordering wavevector is $(\pi/2,\pi/2)$, in contrast to the
$(\pi,0)$ ordering of the FeAs compounds and also the $(\pi,\pi)$
ordering of the insulating parent compound of the cuprate superconductors. ({\em Note} the wavevectors
here are defined with respect to an unfolded zone comprising of a unit cell of one iron atom, oriented along the iron square lattice.  The actual basis vectors are
${\bf \hat{x}}\pm{\bf \hat{y}}$ because of the alternating positions
of the tellurium ions; therefore, crystallographic studies
use a doubled-unit cell, and $a$ and $b$ are along the diagonals
of the iron lattice.)
 The order sets in via a
first order transition at 87$K$, and is accompanied by a monoclinic
distortion\cite{FeTeNeutronPaper}. In the presence of excess iron,
i.e. $Fe_{1+y}Te$, the commensurate order is found to evolve into an
incommensurate spiral. Whether this complex magnetic behavior is
important to understanding other physics in this material is
presently unclear. So far, it is unique to FeTe, where
superconductivity only occurs on substantial alloying with sulfur
$FeS_{0.2}Te_{0.8}$ or selenium $FeSe_{0.5}Te_{0.5}$. However,
besides the interest in understanding the origin of this unusual
magnetism, there are a number of indicators that point to the
presence of strong correlations in FeTe, which would be an important fact to establish in these materials. (1) The
ordered magnetic moment in FeTe is large, $\sim 2\mu_B$, consistent
with a localized $S=1$ at every site. (2) Above the ordering
temperature the magnetic susceptibility falls off in a Curie Weiss
fashion, roughly consistent with the observed ordered moment and
transition temperature. (3) given the absence of Fermi surface
nesting at this wavevector, a spin density wave scenario seems less
favorable than a local moment picture. Furthermore, ARPES
experiments on this material see no obvious nesting at the desired
wave-vector\cite{HasanArpes}, and (4) optical conductivity, which observes a clear SDW
gap in the FeAs materials, does not see an analogous gap in
FeTe\cite{FeTeConductivity}. Another mechanism is therefore
needed to explain the peculiar spin ordering.  Here we will assume that the magnetism in FeTe arises from local magnetic moments on the iron sites, that arise from strong correlations. However, given that FeTe is metallic (with a residual resistivity of $0.2\mathrm{m}\Omega$ cm) - we will have to consider them as being `self-doped'. The key point of this paper is that several of the puzzling magnetic properties of FeTe can be naturally explained if we assume it is near a correlated insulating state with spin and orbital degeneracies. Furthermore, the structural distortion here will be explained as arising, at least partially, from orbital ordering - rather than spin lattice interaction as is usually assumed.  A model
for FeTe's ordering in which the lattice distortion is assumed from the outset is in  \cite{FangOrdertheory}. Other
scenarios based on an itinerant electron picture have also been put
forward \cite{Mazin}, such as the possibility that electron doping
is large enough to change the Fermi surface shape and lead to nesting\cite{HanSavrasov}.

The main assumptions we make in this paper are: {\em(i)} FeTe is
proximate to an $S=1$ magnetic insulator. {\em(ii)}Each site
of this insulator has orbital degeneracy ($d_1,\,d_2$) and
the Jahn Teller effect leads to orbital ordering which
orients the orbitals along the diagonal direction of the
Fe lattice. The orbital $d_1$ accommodating the local moment is uniformly oriented on all sites. {\em (iii)} A small excess density of charge carriers is present in the other Jahn Teller orbital $d_2$.

From these assumptions we show that the  $(\pi/2,\pi/2)$ magnetic
order can be naturally explained, as well as the incommensuration
induced by excess iron. The key mechanisms are the formation of one
dimensional chains induced by the Jahn-Teller ordering, which are
coupled together by double exchange. Part of the structural
transition in this scenario is caused by orbital rather than spin
ordering. Since this proposal invokes an intertwining of the spin, charge and orbital physics, several testable consequences also emerge for transport and ARPES experiments. For example, transport within a single domain is expected to be highly anisotropic, with larger conductivity along the {\em ferromagnetically} ordered diagonal directions. This can potentially be probed by optical conductivity experiments. Indeed, a number of anomalies are already observed in transport properties, albeit in multidomain samples. The electrical conductivity rises sharply below the ordering transition while the Hall effect abruptly changes sign.

Recently, a similar scenario has appeared to
explain the magnetism and structural transition in FeAs \cite{debrink, Singh,phillips}. In contrast to our assumption (ii), the Jahn-Teller
orbitals are assumed to order along the lattice directions. The spin
wave spectrum experimentally observed in FeAs was argued to be much
better described by such a model\cite{Singh}. Here we focus on FeTe,
for the reasons described previously, but our approach is very similar in spirit. If indeed this
mechanism is more general, it allows us to unify phenomena across
this family of compounds.

This paper is organized as follows. We first describe microscopic strongly correlated models for FeTe, which have S=1 and orbital degeneracy. We then consider how orbital ordering of a particular kind can drive magnetism, leading to the observed spin order. The characteristics of spin wave dispersion within this scenario are presented, and the origin of incommensurate magnetism with excess iron, is discussed. Finally, experimental consequences of this scenario for nonmagnetic properties, like conductivity and ARPES, are described.

\section{F\lowercase{e}T\lowercase{e}: A strongly Correlated Viewpoint}
\subsection{Microscopic Model}
\begin{figure}
\psfrag{dxy}{$d_{xy}$}
\psfrag{dx}{$d_{x^2-y^2}$}
\psfrag{dz}{$d_{z^2}$}
\psfrag{dxz}{$d_{Xz},d_{Yz}$}
\includegraphics[width=.45\textwidth]{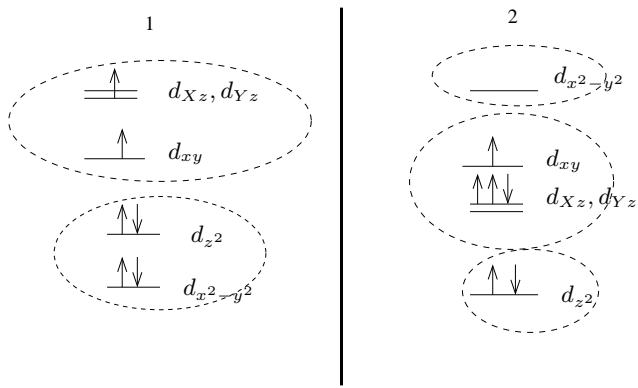}
\caption{\label{Fig1}Two scenarios for the level spacing of the iron atoms' d-orbitals which
lead to orbital degeneracy and $S=1$.
Each set of orbitals grouped together is assumed to fill up according to Hund's
rule before any electrons are added to the next group of orbitals.
This follows from assuming that the crystal field splitting between
orbitals in different groups is greater than the Hund's coupling.
Scenario 1 results in a single electron having to decide between
two degenerate orbitals while Scenario 2 results in a single hole
which is orbitally degenerate. We mainly discuss Scenario 1.}
\end{figure}

We first model FeTe in terms of a nearby correlated insulating
state. The modifications required to account for  metallic
conduction are discussed later. We demand that the insulator carries
net spin $S=1$, per Fe atom, based on the
ordered moment observed at low temperatures. Furthermore, we will
require that they exhibit orbital degeneracy. The only pair of $d$
orbitals that are degenerate in this tetragonal structure are the
$d_{xz}$ and $d_{yz}$ orbitals. Hence we will require an odd filling
of this orbital pair. Two possible microscopic scenarios for the
$d^6$ configuration of the $Fe^{+2}$ ion are sketched in Figure
\ref{Fig1}. In the first, there is one electron available to occupy
the two degenerate orbitals,
while the orbital $d_{xy}$ is singly occupied.
Interestingly, this is the orbital configuration suggested
by the crystal field splitting of the Fe sites. For a perfect
tetrahedral arrangement of $Te$ ions, the $d_{x^2-y^2}$ and
$d_{z^2}$ orbitals lie below the triplet of
$d_{xy},\,d_{xz},\,d_{yz}$. The distortion of the tetrahedron in
this material brings the $d_{xy}$ below the degenerate
$d_{xz},\,d_{yz}$ - leading to the orbital structure in Figure
\ref{Fig1}. Note, the sense of the distortion in FeTe is opposite to
that in FeAs, where a similar exercise leads to a different orbital
ordering \cite{FeTeNeutronPaper}. Since such a local picture of
electronic orbitals may not capture the physics of FeTe we also
point out a different scenario (Figure \ref{Fig1}b), where the
orbital ordering is closer to what is predicted by the band
structure calculations in FeAs \cite{Kuroki}. If we order the
dispersing bands by their center of mass, we end up with the
ordering shown. Here too, the electron assignment can lead to a
$S=1$ orbitally degenerate configuration, but now the degenerate
pair of orbitals contains three electrons. We note that in both
scenarios, the active orbitals are $d_{xy},\,d_{xz},\,d_{yz}$, which
are also the ones expected to be present at the Fermi energy from
weakly correlated band structure calculations of these materials
\cite{Kuroki, WenLee}. The two scenarios are particle hole
conjugates of one another, if one focuses on the active triplet of
orbitals. Henceforth we will assume Scenario 1 for concreteness, as it
corresponds better with experimental facts. The
results there can be easily transcribed to Scenario 2, by a particle
hole transformation.

\begin{figure}
\psfrag{Te}{Te}
\psfrag{A}{$A$}
\psfrag{B}{$B$}
\psfrag{x}{$x$}
\psfrag{X}{$X$}
\psfrag{y}{$y$}
\psfrag{Y}{$Y$}
\psfrag{dxz}{$d_{Xz}$}
\psfrag{dyz}{$d_{Yz}$}
\includegraphics[width=.45\textwidth]{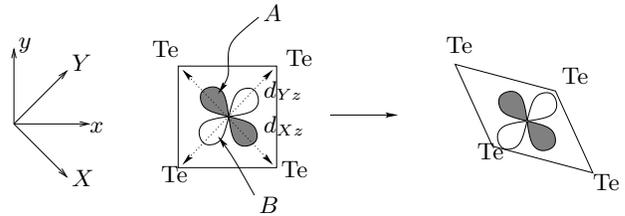}
\caption{\label{fig:JahnTeller}Degeneracy lifting by Jahn-Teller distortion.  The tellurium atoms projected into the
$xy$-plane form a square in the undistorted compound, which distorts into a rhombus, so that $A>B$, lowering the energy of the shaded orbital.}
\end{figure}

\subsection{Orbital and Magnetic Order }
Let us discuss first the the pair of degenerate orbitals in Scenario
1, filled by a single electron. Later we will include the third
orbital. It is convenient to rotate orbitals by 45$^o$, and define
another basis that point along the diagonals
$d_{Xz/Yz}=\frac1{\sqrt{2}}(d_{xz}\pm d_{yz})$. Let
$d^\dagger_1,\,d^\dagger_2$ be the creation operators for electrons
in these diagonal orbitals. The Hamiltonian for this system is: \be
H = H_{KE}+H_U+H_{JT}
 \ee
where the first term is the hopping hamiltonian, and the second and
third terms refer to interactions and coupling to lattice phonons
that lead to the Jahn-Teller effect.

\be H_U = \sum_r \frac{1}{2}U(n_r-1)n_r -J_H\vec{S}_{1r}\cdot\vec{S}_{2r}
\label{SiteU}\ee where $n_r$ is the electron density at site $r$ and
$\vec{S}_ar = \frac12 d^\dagger_{ar}\vec{\sigma}d_{ar}$ is the spin
on site $r$ in orbital $a$. In the limit of strong repulsion $U$ and
a single electron per site, $\langle n \rangle = 1$ we obtain an
insulating state with orbital degeneracy. This degeneracy is
typically resolved by the Jahn-Teller effect. A lattice distortion
which breaks symmetry and splits the degeneracy occurs. The precise
distortion that is realized is hard to predict, so we will assume that
it is indeed of the type required to obtain the structural
transition seen in this material. This involves a uniform
orthorhombic distortion that changes the relative lengths of the two diagonal bonds. If we denote by $A$ and $B$, the classical bond
lengths for the two diagonals (so $A=B$ in the tetragonal state),
then we will assume
that the lattice coupling is given by: \begin{equation} H_{JT} =
-\alpha(A-B)\sum_r(d^\dagger_{1r}d_{1r} -
d^\dagger_{2r}d_{2r})+\frac{\beta}{2}(A-B)^2. \end{equation}
(See Fig. \ref{fig:JahnTeller}.)
The
orbital ordering implies that, in scenario 1, the single electron
on each site always occupies the same orbital, i.e.,
$n_{1r}=1$ and $n_{2r}=0$, or vice versa.
For concreteness let us
suppose that the $X$ diagonal expands.
  Then
the electrons occupy orbital $1$
($d_{Xz}$) while orbital $2$ ($d_{Yz}$)is empty, because it is higher
in energy.
Since each site has an unpaired electron, we can now derive the
Hamiltonian governing their magnetic moments.

In the insulating limit, the magnetic interaction is generated by
virtual hopping of electrons. To proceed, we need to specify
$H_{KE}$. Clearly, given the geometry of the $Xz, /,Yz$ orbitals,
hopping along the diagonals will be very anisotropic. We denote
the $X$ ($Y$) diagonal hopping of the $d_{Xz}$ ($d_{Yz}$) orbital by
$t_2$, and let $t_2'$ be the
hopping of each
in the orthogonal direction (see Fig. \ref{fig:weavingtogether}).
The figure suggests that
$t_2 \gg t_2'$ (see \cite{2band}), and for FeAs the nearest
neighbor hoppings are comparatively small as well.
Hence, we will simply work with the exchange interaction induced by
$t_2$. Note, the next neighbor hopping only operates within a single
sublattice (labeled $A$ and $B$ in the figure), so here we consider
just the $A$ sublattice. The $t_2$ hopping will introduce
antiferromagnetic exchange, but only along the $X$ diagonal
$J_{2X}=t_2^2/U$. This will lead to antiferromagnetic
order along this diagonal direction. Note, assuming Scenario 1, this
antiferromagnetic direction will be the expanded diagonal of
the distorted compound, which is consistent with the observed
wavenumbers of the magnetic and structural distortions\cite{FeTeNeutronPaper}.
Note, although
the diagonal chains are ordered, there is negligible coupling
between chains at this moment (because of the smallness of $t_2'$).
Below, we will see there is a more important mechanism that can lock
the magnetic order in the chains together - the double exchange
interaction.

\subsection{Double Exchange}
Hopping  of electrons between the antiferromagnetic chains can
readily occur if they occupy the $d_{Yz}$ orbital. However, in a
orbitally ordered insulator, these are assumed to be empty. Given
the empirical fact that FeTe is a metal, we assume a small
occupation $\delta$ of electrons in this orbital. This can arise
because of `self-doping', i.e. if the orbital dispersion cannot be
completely neglected, a fraction of carriers from one of the
`filled' bands (for example $d_{xy}$), could be transferred to an
empty band, i.e. this orbital. This also leads to a conducting state
at stoichiometry, as demanded by experimental observations.
Furthermore, FeTe always occurs with a slight excess of Fe i.e.
Fe$_{1+y}$Te. The electrons from the
additional Fe$^{+2}$ ions also contribute to $\delta$.

\begin{figure*}
\psfrag{t2}{$t_2$}
\psfrag{t2'}{$t_2'$}
\psfrag{J2X}{$J_{2X}$}
\psfrag{t2d}{$t_2\delta$}
\psfrag{x}{$x$}
\psfrag{y}{$y$}
\psfrag{X}{$X$}
\psfrag{t2}{$t_2$}
\includegraphics[width=\textwidth]{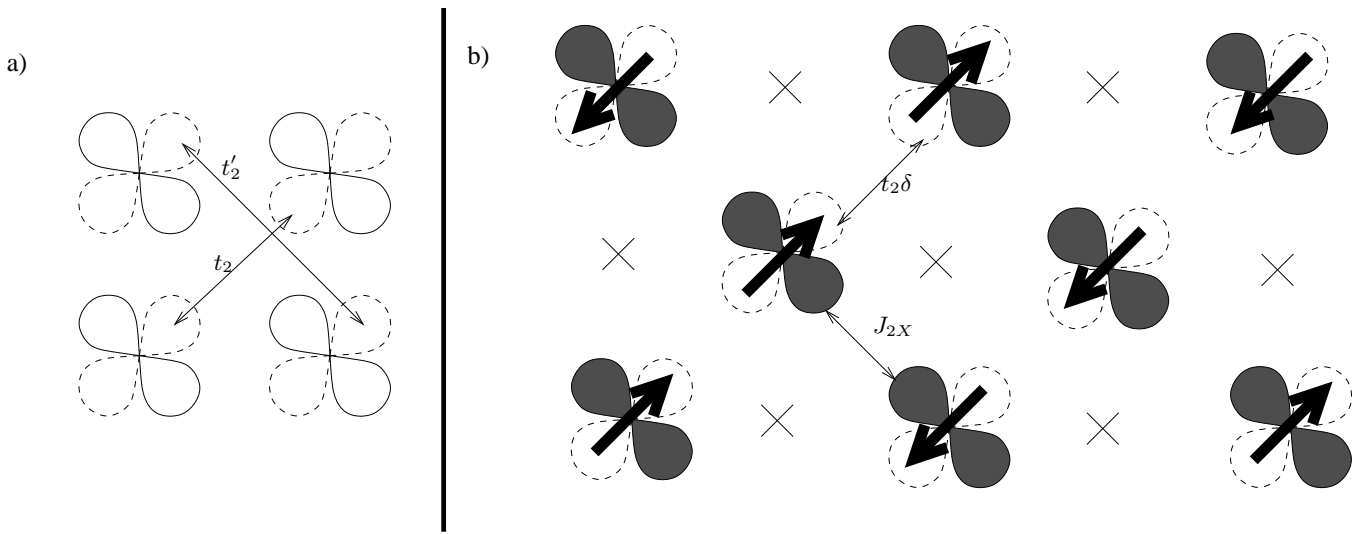}
\caption{\label{fig:weavingtogether}
How the spin-order on sublattices arises out
of the orbital order. The solid and
dotted lines show the projections of the $d_{Xz}$ and $d_{Yz}$ orbitals. a) The
second nearest neighbor hopping amplitudes, which
determine the order in the sublattices. Electrons stay in the same orbital
when they hop to the second nearest neighbor.  The hopping for each
orbital is anisotropic,
with an anisotropy that depends on the orbital, so each orbital forms
a set of one-dimensional chains.  Electrons
can hop between nearest neighbors as well (not shown).  b) Generating
the spin order
in the A-sublattice.  (The iron atoms in the B-sublattice are indicated
by x's.) The $d_{Xz}$ are shaded to indicate that
they are filled and the $d_{Yz}$ orbitals are lightly doped.
The $d_{Xz}$ orbital form one-dimensional Mott
insulators parallel to the $X$-axis
with
antiferromagnetic coupling $J_{2X}$.  The $d_{Yz}$ orbitals form
one-dimensional metals parallel to the $Y$ direction. Ferromagnetic
order along $Y$ lowers the kinetic energy of the metals
by about $t_2\delta$.
}
\end{figure*}

Such an excess carrier density will lead to ferromagnetic
interactions between chains, via the double exchange mechanism.  An
electron hopping in this nearly empty orbital will be Hund's coupled
to the $d_{Xz}$ electron according to Equation \ref{SiteU}. The
Hund's coupling is typically large and will force both electrons to
have the same spin. If the electron is to hop to the neighboring
chain,  the spins must be parallel. Then, it can enjoy a lowering of
kinetic energy by $-2t_2$. Thus, a ferromagnetic arrangement of
spins will have a lower energy than an antiferromagnetic arrangement
by an amount $2 |t_2|\delta$, which can roughly be viewed as a
ferromagnetic coupling along the $Y$ diagonal (strictly speaking
this is a nonlocal interaction, and cannot be assigned solely to the
diagonal bond). Thus $J_{2Y}=-2|t_2|\delta$. With this combination
of exchange constants, the magnetic ordering on a single sublattice
is shown in Figure \ref{fig:weavingtogether}. Note that it has the
wavevector $(\pi/2,\pi/2)$, as required.

\subsection{Coupling the Sublattices}
So far, the two sublattices ($A$ and $B$) are independent.  When
the doping is small,
the dominant spin interaction between nearest neighbor sites will
arise from antiferromagnetic
exchange from the electrons in the half filled $d_{xy}$
orbital.  (The nearest neighbor hopping of electrons in the $d_{Xz}$
orbitals would induce a smaller ferromagnetic interaction
controlled by the Hund's coupling,
since the dominant hopping is expected to switch the orbital index.)
We will also invoke a coupling
to the lattice to generate a biquadratic interaction term in order
to lock in the commensurate wave number, leading to the net Hamiltonian
describing the interaction of spin 1 atoms on a square lattice:
\begin{equation}
H = \sum_{ij} J_{ij}{\bf S}_i\cdot {\bf S}_j - K \sum_{\langle ij \rangle} ({\bf S}_i\cdot {\bf S}_j)^2
\label{SpinModel}
\end{equation}
$J_{2X} = t_2^2/U$, $J_{2Y}=-2|t_2|\delta$ and $J_1>0$, and
$J_1$ is the same for both nearest neighbor bonds. The phase diagram of this model as a function of increasing $K$ is included on the $y$-axis of Fig. \ref{fig:phasedi}. The ${\bf S}_i\cdot{\bf S}_j$ terms alone would lead to an incommensurate spiral state \cite{Xu}, where the nearest neighbor spins in the spiral state are close to being  orthogonal to one another (the
``single spiral state" below).
This state, even in the small incommensuration limit, is significantly different from the experimentally observed collinear state.

To stabilize the commensurate $(\frac{\pi}{2},-\frac{\pi}{2})$ state,
we take into account the
biquadratic spin interaction ($-K({\bf S}_1 \cdot {\bf S}_2)^2$)
which is a well-known consequence of spin-lattice coupling.
This term prefers collinear magnetism.
As we will see below, even  modest values of the spin phonon coupling $K$ can induce locking of the commensurate, collinear phase observed in experiments. In particular, we show below that the critical coupling required to induce collinear order $K\gtrsim J_1^2/J_{2X}$ can be parameterically smaller than $J_1$. Coupling of spin and lattice is presumably essential to getting a commensurate state at this wavevector.

{\em Phase Diagram:} The phase diagram can be obtained by taking the
ansatz of a pair of coplanar spirals on the two sublattices with an
arbitrary phase $\phi$ between them, and wavevector $k$ along diagonal $X$.
This is probably sufficiently general to
capture the ground states of Eqn. \ref{SpinModel}, because
the ferromagnetic coupling along the $Y$-direction prevents the spin
from varying in that direction. Let
the azimuthal angle of spin $i$ be $\theta_i$. Then the ansatz reads:
\begin{equation}
\theta_i=k(x_i-y_i)+(-1)^{x_i+y_i}\frac{\phi}{2}.
\label{eq:rhythm}
\end{equation}
Here,$(k,-k)$ is the wave number of the spin arrangement.
(This state is ICA, from Ref. \cite{FangOrdertheory}.)
Note, the
experimentally observed collinear state corresponds to $k=\pi/2$ and
$\phi=\pi/2$. Along any row, the angle between adjacent
spins alternates between $k-\phi$ and $k+\phi$. For $k=\phi=\pi/2$, the
spins therefore alternate from parallel to antiparallel.

Minimizing the energy per site,
\begin{alignat}{1}
E[k,\phi] = J_{2X}\cos 2k+2&J_1\cos k\cos\phi\nonumber\\
&-K(1+\cos 2k\cos 2\phi)-2|t_2|\delta
\label{eq:Evarl}
\end{alignat}
over $\phi$ and $k$ shows that there is a
transition between a pure spiral with $\phi=0$ and incommensurate
wavevector $k=\frac{\pi}{2}+O(\frac{J_1}{J_{2X}})$  and the
collinear state $k=\pi/2,\,\phi=\pi/2$ when $4K(J_{2X}-K)=J_1^2$.
For small $K\ll J_{2X}$, the critical $K$ is $\approx
J_1^2/{4J_{2X}}$.

We can understand this result intuitively as follows. The spiral
order occurs to lower the nearest neighbor energy.  The coupling of
a spin to its neighbors to the left and right cancels in the
perfectly collinear state.  To take advantage of $J_1$, the
neighbors should therefore make an angle less than $180^{\circ}$,
and the central spin should point opposite to the sum of the two
neighboring spins.  If $J_1$ is small compared to $J_{2X}$, then the
neighbors are \emph{nearly} antiparallel, so $J_1$ has a very weak
effect, explaining why the critical value for $K$ is not of order
$J_1$ as one might have expected, but rather second order in $J_1$.

\begin{figure}
\psfrag{kX}{$k_X$}
\psfrag{kY}{$k_Y$}
\psfrag{om}{$\omega$}
\includegraphics[width=.45\textwidth]{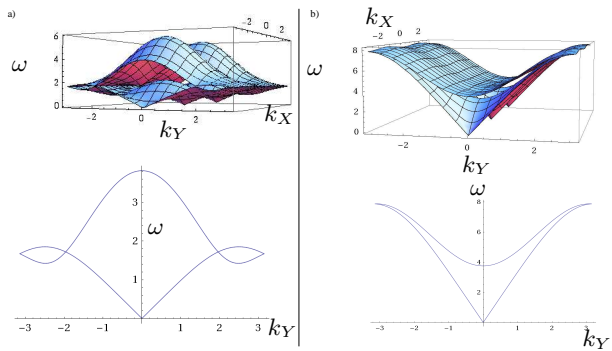}
\caption{\label{fig:bubblefilm}
Comparison of the spin wave dispersion for order stabilized by $J_3$ versus
by a ferromagnetic diagonal interaction. The dispersions
are plotted as a function of $(k_X,k_Y)=(k_x-k_y,k_x+k_y)$ in the upper
pictures.  The other graphs show the dispersion in the $Y$ direction.
a) The
spin wave spectrum for $J_{2X}=J_{2Y}=1, J_1=1.47,K=0.265,J_3=0.4$,
which have the same effective $J_{1w,s}$ as in Ref.\cite{FangOrdertheory}.
b) This shows the spectrum for $J_{2X}=-J_{2Y}=1,J_1=0.7,K=0.265,J_3=0$.
The sign of $J_{2Y}$ is flipped to describe a ferromagnetic coupling,
while $J_3$ is set to zero.  ($J_1$'s value is decreased relative
to a) to stabilize the order.)
Note that the upper band changes from a hill shape
to a valley.
}
\end{figure}

{\em Spin Waves:} To contrast this scenario with others that predict
the same magnetic ordering pattern, we calculate the spin wave
spectrum for the model \ref{SpinModel}. We expect inelastic neutron
scattering experiments in the future to be able to check this
prediction. In particular, we contrast it with a recent theory\cite{FangOrdertheory,OldHu}
in which there is a sufficiently large third neighbor exchange $J_3$ and
the monoclinic distortion alters the
first and second nearest neighbor interactions, leading to
the observed magnetic
order.  In our model, the magnetic order is stabilized just by the
anisotropy of $J_2$, with one ferromagnetic and one
antiferromagnetic direction.  (Ref. \cite{FangOrdertheory,OldHu} also includes
the
anisotropic couplings but the third neighbor
hopping makes it possible to stabilize the order without
ferromagnetic directions.)
Note, while doing the spin wave calculations, we expand about the
equilibrium spin state, and hence the biquadratic interaction
effectively leads to weak ($w$) and strong ($s$) nearest neighbor
bonds, $J_{1s} \neq J_{1w}$. Thus, our spin wave dispersion is
reproduced by a model involving only quadratic spin couplings, where
$ J_1{\bf S}_1\cdot{\bf S}_2-K_1({\bf S}_1\cdot{\bf S}_2)^2
\Rightarrow (J_1-2K_1\langle {\bf S}_1\cdot{\bf S}_2\rangle){\bf
S}_1\cdot{\bf S}_2$ Therefore the bonds between parallel spins are
effectively weaker than those between antiparallel spins, similar to
Ref.  \cite{FangOrdertheory,OldHu}.
The spin wave spectrum is obtained using the Holstein-Primakoff expansion (see e.g. \cite{Madelung}).  Figure \ref{fig:bubblefilm} compares the dispersions one expects in the two models, the first  with a strong $J_3$ and ours with a ferromagnetic diagonal coupling. Note, the upper band of the dispersions curves
in the opposite direction along the $Y$-axis, because of the ferromagnetic 
coupling $J_b$.

\subsection{Doping Induced Incommensuration}
The properties of $Fe_{1+y}Te$ have been experimentally
investigated as $y$ is varied.  Experimentally, $Fe_{1+y}Te$ is
found to have incommensurate spin order when $y$ is large enough,
with an incommensurate wavevector that deviates from $(\pi/2,\pi/2)$
linearly with doping \cite{FeTeNeutronPaper}. One of the effects of the excess iron, which is believed to be in the $Fe^{2+}$ state, is to electron dope the system by $2y$ electrons.Here we consider
how this may be explained as a result of the increasing electron density $\delta=2y$ in the $Yz$ orbital, which makes
nearest neighbor hopping more important. This will lead to incommensurate order that is related to the doping level.

To see this we will consider two dimensional motion of
the doped electrons, which can occur only if
 hopping amplitudes other than $t_2$ are taken into account.
These types of motion are limited by the wide stripes of the commensurate
$(\frac{\pi}{2},\frac{\pi}{2})$ spin pattern, increasing the kinetic
energy, according to the uncertainty principle.
Such kinetic
energy effect has a stronger dependence on
the relative angle between a pair of spins when the
angle is close to $90^{\circ}$ than the interactions considered
above,
so it is able to distort the commensurate collinear state with wave
vector $(\frac{\pi}{2},-\frac{\pi}{2})$ and $\phi=\pi/2$, into a {\em
``double spiral"}, a slightly twisted version of the same state,
of the form Eq. (\ref{eq:rhythm}), and with
$\phi\approx\frac{\pi}{2}$ and $k\approx\frac{\pi}{2}$, even when
the doping is weak.  

\begin{figure}
\psfrag{0.1}{0.1}
\psfrag{0.2}{0.2}
\psfrag{0.3}{0.3}
\psfrag{0.3}{0.3}
\psfrag{0.4}{0.4}
\psfrag{0.5}{0.5}
\psfrag{K/J2X}{$\frac{K}{J_{2X}}$}
\includegraphics[width=.45\textwidth]{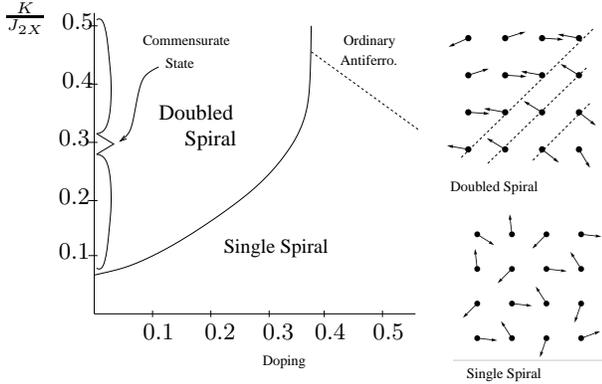}
\caption{\label{fig:phasedi}
The phase diagram as a function of $K$ and doping for
$J_1=\frac{1}{2}J_{2X}$, $t_2'=1.5J_{2X}$, using the infinite Hund's
coupling approximation. The three phases
are the doubled spiral, the single spiral, and ordinary $(\pi,\,\pi)$ antiferromagnetism.
Note that both
the single and doubled spirals are constant along wavefronts
parallel to $Y$. In the perpendicular direction, the double spiral alternates between rotating through small angle and angles close to $180^\circ$, while the single spiral rotates through around $90^{\circ}$  each time. 
A calculation which takes fluctuations or the finiteness of the
Hund's constant into account would find a finite range of dopings where
the commensurate phase is stable.
The solid/dashed
line represents a discontinuous/continuous transition. The discontinuous
transition boundary should actually be replaced by a wide swath of
phase separation.
}
\end{figure}

To understand the kinetic energy effect, we assume a \emph{uniformly}
varying classical spin
configuration and minimize the energy of a single Hund's coupled
electron, Eq. (\ref{SiteU}),
hopping in this background. The resulting kinetic energy
($KE_{min}$) times electron density, $KE_{min}\delta$ is added to the
magnetic energy (\ref{SpinModel}), and this total energy is
minimized to obtain the phase diagram in Fig. \ref{fig:phasedi}.
A shortcoming of this phase diagram
is that the spiral phases may not be stable against phase 
separation (see below).

We will now focus for simplicity on determining the kinetic energy due to
hopping via $t_2'$. We checked also that the
nearest neighbor $t_1$ of Ref. \cite{2band} leads
to a very similar phase diagram.  The kinetic
energy is computed
from the hopping Hamiltonian
$H_{t_2'}=-t_2'\sum_{i}d^\dagger_{Yz}({\bf r}_i+{\bf \hat{x}}-{\bf
\hat{y}})d_{Yz}({\bf r}_i)+h.c.$, which favors ferromagnetic alignment
along the $X$-diagonal.Since the Hund's energy is larger than the $t$'s, the natural basis for
spin states on each site are the states quantized along the local spin
orientation of the magnet ($\theta_i$).
If the Hund's
energy is assumed to be infinite,
we can assume that the electron state is always aligned with
the local spin.  Such a state, with momentum ${\bf p}$
is represented by the
following electron wavefunction,
\begin{equation}
\psi_i =
\frac{1}{\sqrt{2}}\left(\begin{array}{c}e^{-i\frac{\theta_i}{2}}\\e^{i\frac{\theta_i}{2}}\end{array}\right)e^{i{\bf
p}\cdot{\bf r}_i}.
\end{equation}
We can now find the variational energy of this wave function, and minimize with respect to $p$.

 The
expectation value of the kinetic energy for a bond between a pair of
sites is $2t\Re(\psi_i^{\dagger}\psi_j)$. Aside from the phase
factor from the momentum, this overlap is proportional to $\cos k$
for spins adjacent along the $X$ direction (the cosine of half the
angle between the spins). In particular,
because the spin is conserved during the
hopping, the electron cannot hop at all onto a site
with an antiparallel spin. The electron kinetic energy is
$2t_2\cos (p_x+p_y)+2t_2'\cos(\frac{p_x-p_y}{2})\cos{k}$. Minimizing
over $p$, we obtain $KE_{min}=-2|t_2|-2|t_2'\cos(k)|$.
 Combining this
with the previously obtained energy in the absence of $t_2'$, Eqn.
(\ref{eq:Evarl}), and minimizing with respect to $\phi$ and $k$ we
obtain the double spiral.
This can be easily seen when the doping is very weak: focusing on
the competing terms $J_{2X}$ and $t_2'\delta$, and assuming
$k\approx\frac{\pi}{2}$ to find $E=J_{2X}\cos(2k)-2t_2'|\cos
k|\approx -J_{2X}+2J_{2X}
(k-\frac{\pi}{2})^2-2t_2'\delta|k-\frac{\pi}{2}|$, so
$|k-\frac{\pi}{2}|=\frac{t_2'\delta}{2J_{2X}}$ minimizes the energy.
The incommensuration therefore is proportional to the doping
strength, as observed experimentally.


If we allow the electrons to hop virtually into states that
violate Hund's rule, the commensurate state survives over
a finite range of dopings.  When the twisting is very small, it becomes easier
for an electron that is trying to hop along the antiferromagnetic
diagonal to hop through the Hund's violating states than to
stay in states that are parallel to the local spin but
which have very small overlaps.  In this case, there is no
reason for the spin order to distort at all.  The virtual
hopping has the greater efficiency
when $|t_2'\cos k|<|\frac{t_2'^2}{J_H}|$, or
(using the result for $k$ at small $\delta$),
$\delta\lesssim\frac{J_{2X}}{J_H}$. (A more
detailed calculation gives the same result.) Quantum fluctuations also allow
the electrons to hop more easily in this direction, by temporarily
making adjacent atoms become oriented parallel to one another.

A very similar phase
diagram appears for $t_1$ hopping.  This hopping
probably has weaker effects, however, since
the orbital switches with each hop. This implies that an electron
alternates between Hund's rule violating and satisfying
states, so that the amplitude for a pair of these steps is of
order $\frac{t_1^2}{J_H}$.
\emph{This} process
leads to an antiferromagnetic
interaction along rows and columns, but this causes incommensuration
just as the ferromagnetism induced by motion along the diagonals did.

Although these calculations explain how incommensurate order can
occur, the order differs from the experimentally proposed pattern in \cite{FeTeNeutronPaper}. The proposed order has the spin dominantly along the $\hat{Y}$ direction whose magnitude is modulated with an incommensurate wavevector, and also a spin spiral composed of the orthogonal spin directions, at the same wavevector.  
While further experimental work is required to confirm the true nature of the complex incommensurate order, we note one interesting measurement, 
that orthorhombic
symmetry is recovered at higher doping, (eg. at $x=0.141$ \cite{FeTeNeutronPaper}) where the
incommensurate state is stabilized.  This is consistent with our
assumption that the monoclinic part of the distortion is strongly
coupled to spin order. Fig. \ref{fig:rhombi} shows
that the sense of the monoclinic distortion is
correlated with the bond energies $\vec{S}_i\cdot \vec{S}_j$. 
Shifting the spin ordering and thus the pattern of bond energies
over one site would cause the lattice to tilt in the opposite direction along the $Y$ axis.
For an incommensurate order of the kind proposed in Ref. \cite{FeTeNeutronPaper},
these bond energies are also modulated with an incommensurate
wavevector which removes the monoclinic distortion.
In contrast, for the orders described by Eqn.\ref{eq:rhythm}, including both the double and single spirals
we have considered, the bond energies are independent of
incommensuration, hence a monoclinic distortion is expected
throughout. An example of a spiral order that would not induce a monoclinic distortion, close to the commensurate state of interest  is: $\theta_i=(-1)^{x_i+y_i}k(x_i-y_i)$, with $k$ close to $\frac{\pi}{2}$.
Effects that we have neglected, including spin anisotropy and the effect of the excess iron moments, can modify the precise form of the incommensurate state.

\emph{Phase Separation} An additional shortcoming of this explanation of
the incommensurate order is that the spiral
phases can be unstable to phase separation\cite{Dagotto}.  
The doped electrons prefer to
be segregated into high and low density regions, with
different spin orders.  The predictions above can still
be relevant, because repulsive
forces between electrons help to limit the phase separation.

\begin{figure}
\includegraphics[width=.45\textwidth]{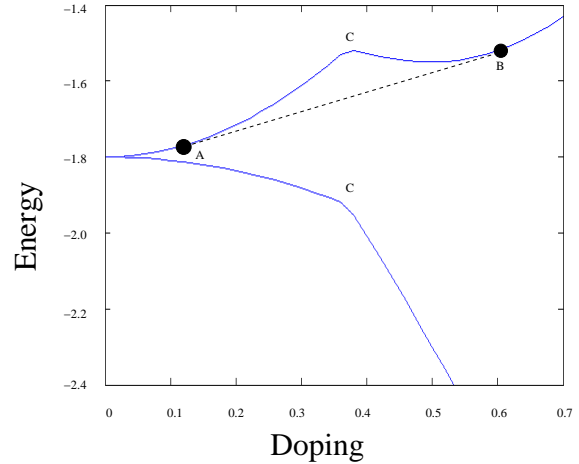}
\caption{\label{fig:helmholtzcons}
Energetics of phase separation. The solid curve shows the energy of states
with uniform doping, with repulsion $V=6J_{2X}$ (upper 
curve) and without. The parameters are the same
as in Fig. \ref{fig:phasedi}, with $K=0.4$. 
The concave down portion of each energy
curve is unstable to phase separation. 
The points labelled
$C$ show the unphysical transition between the doubled- and single-spiral
states. The dashed line corrects the upper
energy-curve, by showing the energy
of the state where the $A$ and $B$ phases coexist.}
\end{figure}

Fig. \ref{fig:helmholtzcons} 
shows the energy as a function of doping both without
and with a strong \emph{short-range} repulsion, described by adding
a term $\frac{1}{2}V\delta^2$ to the energy functional. In the figure,
$K=0.4$ and the other parameter values taken from Fig. \ref{fig:phasedi}.
A concave down portion is unstable to phase separation.
For a doping in between $A$ and $B$, i.e., $0.1\leq\delta\leq 0.6$,
dividing the compound up into two
regions with the dopings $0.1$ and $0.6$ gives a
state with a lower energy, whose energy is represented by the tangent.
In this case the phase mixture is between a doubled spiral
with a low electron density and an antiferromagnetically
ordered high density portion. 

The Coulomb interactions can have one of two
consequences; they either stabilize a uniform
state as just described (if the force is strong
enough at short distances), or else they force
the two coexisting phases to fill small alternating
portions of the FeTe, 
rather than becoming completely segregated.  
The ordering with alternating regions of 
the $(\frac{\pi}{2},\frac{\pi}{2})$ and $(\pi,\pi)$ phases
could also be incommensurate.

{\em Additional Doping} A surprising order may arise if the compound is
sufficiently doped, which directly connects to our central
assumption that the orthorhombic part of the structural distortion
arises from orbital, rather than spin ordering.
In our model, large electron doping would drive
ferromagnetic order on each of the A and B sublattices.
This optimizes the electron kinetic energy. The sublattices
continue to couple together antiferromagnetically,
leading to $(\pi,\pi)$ magnetic order.
In Fig. \ref{fig:phasedi},
the order varies from commensurate, to the rhythmic spiral phase to
the $(\pi,\pi)$ antiferromagnetic phase as the doping increases
along a line at $K=0.5J_{2X}$.
For
a smaller value of $K$, the pure spiral phase occurs before
$(\pi,\pi)$ antiferromagnetism sets in. Note however, that while
$(\pi,\pi)$ antiferromagnetic order is compatible with tetragonal
symmetry, here we expect orbital order to persist and result in an
orthorhombic distortion, clearly signaling the independence of
structural and magnetic order.

On the other hand, {\em hole} doping weakens the kinetic energy
effect to the point where each sublattice has ordinary
antiferromagnetic order, and the biquadratic interaction orients the two sublattice-spins parallel to each other, leading to $(\pi,0)$
magnetic order of the type seen in the FeAs materials.
Again, while the $(\pi,0)$ ordering would be accompanied by
compression along the $x$ or $y$ axis, if this were magnetically
driven, here we expect the orbital ordering to remain the same, so
the diagonal distortion would not change.   A more direct test of whether
the distortion or the magnetic ordering
is the primary phenomenon is to apply a strong magnetic field
to eliminate the magnetic ordering and see whether the distortion remains.
This would be possible if the exchange interactions are weak, or can
be weakened by modifying the compound somehow, e.g., by applying pressure.

\section{Transport and Single Electron Properties}
The main consequence for transport of the orbital ordering induced
quasi one dimensionality is that the conductivity should be strongly
anisotropic. Below the orbital and magnetic ordering temperature,
the excess electrons in the $Yz$ band move much more readily in the
$Y$ direction, because $t_2$ is greater than $t_1,t_2'$.
Furthermore, the spins are
ferromagnetically aligned along this direction, and can hence
propagate easily.  To travel in the orthogonal direction, they must
cross through diagonals where the spins are oppositely aligned.
While a difference in electrical conductivity
along the two diagonal directions is hardly surprising given the
symmetry of the low temperature state, the specific prediction here
is that this will be a significant effect, and the nature of the
anisotropy is that the low conductance is to be found along the
antiferromagnetically ordered diagonals.

Experimentally, a Drude peak has been observed to develop below the
ordering transition \cite{FeTeConductivity}. Since a test of
anisotropy demands a single domain, optical conductivity on a sample
where the domain size is larger than the spot size is required, and
should be feasible. There, depending on the direction of the
polarization, a different conductivity should result. Note, writing
$\sigma(\omega) = \frac{ne^2\tau}{m^*} \frac{1}{1+i\omega \tau}$,
both the effective mass $m^*$ and the scattering rate $\tau$ are
expected to be anisotropic, which can be separately inferred from a
knowledge of $\sigma(\omega)$.The anisotropy in this scenario is
expected to be particularly prominent in the low frequency limit,
since scattering by spins will limit the scattering time along the
antiferromagnetic diagonal.  At higher frequencies, only the anisotropy
of $m^*$ will have an effect.  However,
if the scattering rate for the $X$ direction
is extremely large, $\sigma(\omega)$ will be flat in that
direction, while conductivity along $Y$ will have a Drude peak.



\begin{figure}
\psfrag{S}{S} 
\psfrag{kX}{$k_X$}
\psfrag{kY}{$k_Y$}
\includegraphics[width=.3\textwidth]{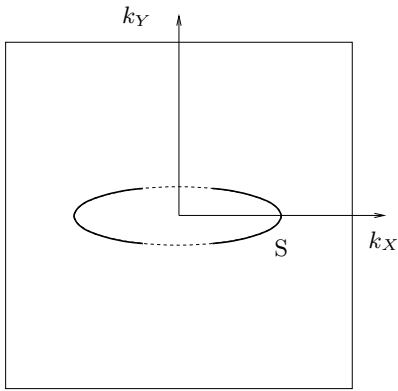}
\caption{\label{fig:sandp} The shape of the Fermi surface of the
$Yz$ electrons in the actual Brillouin zone.  We predict that the
surface is narrow in the ferromagnetic direction.  Furthermore,
certain parts of the ellipse should
disappear in ARPES with polarized light. The
dashed portions of the Fermi surface fade out for $S$ polarization.
This tests the assumption that the
anisotropy results not from the lattice distortion but because the 
dispersing electrons are in a single
orbital.  }
\end{figure}

{\em ARPES} Given the quasi one-dimensional dispersion along the
diagonals, a narrow elliptical Fermi surface tilted at $45^{\circ}$
to the Fe-Fe bonds, is expected to appear below the ordering
temperature. Moreover, these would be orbitally polarized, which can
be experimentally tested using polarized light to determine orbital
content along high symmetry directions. When the scattering plane containing the photon and the ejected electron is perpendicular to the surface, and along, say, the $Xz$ plane, selection rules imply that
$S$ ($P$) polarized light with polarization perpendicular to (parallel to) the scattering plane, ejects only electrons in the $Yz$ ($Xz$) orbital. The strongly dispersing direction should therefore appear for $S$ polarized light, 
and will have the form shown in Figure \ref{fig:sandp}, 
if a single domain is imaged. 
 The polarization dependence provides an
experimental signature even from multi-domain data.  Current ARPES data on FeTe \cite{HasanArpes} has
not reported such a signature, however, the
intensity associated with this Fermi surface segment is hard to
estimate reliably. Direct experimental probes of orbital ordering should also 
be able to test this scenario. We hope this prediction will stimulate further experiments.


\section{Conclusions}
We have argued that FeTe, a material closely related to the recently
discovered Fe based superconductors, is likely to be a fairly
strongly correlated material. This motivates us to use a local
picture of the electronic structure, which in turn can explain,
quite naturally, the unusual magnetism observed in this material.
The key ingredient of this scenario is an emergent quasi one
dimensionality arising from orbital ordering along the diagonals.
While this theoretical scenario is a simplified
caricature of the real system, it does make several qualitative
predictions for experiments which should be readily
testable.  The conductivity should be anisotropic,
with higher conductivity along the ferromagnetic
direction, if the unusual magnetic order is caused by
the kinetic energy of conduction electrons.  Certain
 parts of the electron Fermi surface should disappear
in polarized ARPES, indicating the orbital ordering.  Lastly, if
the Jahn-Teller effect causes the orthorhombic distortion, then 
the distortion should persist
even when the magnetic order is removed or changed to one which
would not be expected to favor lattice distortion on the basis
of symmetry.

We thank Eugene Demler for thoughtful conversations,
and acknowledge support from LBNL DOE-504108.

\bibliography{bibli}

\end{document}